\documentclass[12pt]{article}
\usepackage{epsfig,amsfonts,amsmath,array,mathrsfs}
\newcommand{\pbs}[1]{\let\temp=\\#1\let\\=\temp}
\renewcommand{\theequation}{\thesection.\arabic{equation}}
\def\be{\begin{equation}}\def\ee{\end{equation}}
\def\cvp{\raise 2pt\hbox{,}} 

\def\Vect{\mathop{\text{Vect}}\nolimits}

 \def\Tr{\mathop{\rm Tr}\nolimits}

\def\re{\mathop{\rm Re}\nolimits} 
\def\diag{\mathop{\rm diag}\nolimits} 
\def\rank{\mathop{\text{rank}}\nolimits}

\def\d{{\rm d}}
\def\nn{{\cal N}}

 \def\uN{{\rm U}(N)} 
  
\def\La{\Lambda}
\def\wmic{W_{\text{mic}}}\def\wl{W_{\text{low}}}
\def\wsc{W_{\text{SC}}}
\def\wscN{W_{\text{SC}}^{(N)}}
\def\a{\mathbf a}\def\g{\mathbf g}\def\x{\mathbf x} 

\def\uR{\text{U}(1)_{\text R}} \def\u{\text{U}(1)}
\def\lsw{\lambda_{\text{SW}}}

\def\vevab#1{\bigl\langle\a\big|#1\big|\a\bigr\rangle}
\def\plb#1#2#3{{\it Phys.\ Lett.\ }{\bf B #1} (#2) #3}
\def\npb#1#2#3{{\it Nucl.\ Phys.\ }{\bf B #1} (#2) #3}

\def\jhep#1#2#3{{\it J. High Energy Phys.\ }{\bf #1} (#2) #3}
\def\prd#1#2#3{{\it Phys.\ Rev.\ }{\bf D #1} (#2) #3}

\def\atmp#1#2#3{{\it Adv.\ Theor.\ Math.\ Phys.\ }{\bf #1} (#2) #3}
\def\cmp#1#2#3{{\it Comm.\ Math.\ Phys.\ }{\bf #1} (#2) #3}
\def\pr#1#2#3{{\it Phys.\ Rep.\ }{\bf #1} (#2) #3}
\def\jmp#1#2#3{{\it J.\ Math.\ Phys.\ }{\bf #1} (#2) #3}

\begin{document}
%
%
\pagestyle{empty}
{\parskip 0in

\hfill LPTENS-07/33

\hfill arXiv:0707.3885 [hep-th]}

\vfill
\begin{center}
{\LARGE Microscopic quantum superpotential}

\medskip

{\LARGE in $\nn=1$ gauge theories}

\vspace{0.4in}

Frank F{\scshape errari}
\\
\medskip
{\it Service de Physique Th\'eorique et Math\'ematique\\
Universit\'e Libre de Bruxelles and International Solvay Institutes\\
Campus de la Plaine, CP 231, B-1050 Bruxelles, Belgique
}\\
\smallskip
{\tt frank.ferrari@ulb.ac.be}
\end{center}
\vfill\noindent

We consider the $\nn=1$ super Yang-Mills theory with gauge group
$\uN$, adjoint chiral multiplet $X$ and tree-level superpotential $\Tr
W(X)$. We compute the quantum effective superpotential $\wmic$ as a
function of arbitrary off-shell boundary conditions at infinity for
the scalar field $X$. This effective superpotential has a remarkable
property: its critical points are in one-to-one correspondence with
the full set of quantum vacua of the theory, providing in particular a
unified picture of solutions with different ranks for the low energy
gauge group. In this sense, $\wmic$ is a good microscopic effective
quantum superpotential for the theory. This property is not shared by
other quantum effective superpotentials commonly used in the
literature, like in the strong coupling approach or the glueball
superpotentials. The result of this paper is a first step in extending
Nekrasov's microscopic derivation of the Seiberg-Witten solution of
$\nn=2$ super Yang-Mills theories to the realm of $\nn=1$ gauge
theories.

\vfill

\medskip
%
\begin{flushleft}
\today
\end{flushleft}
\newpage\pagestyle{plain}
\baselineskip 16pt
\setcounter{footnote}{0}

\section{Introduction and motivations}
\setcounter{equation}{0}

Since the advent of exact non-perturbative results in four dimensional
supersymmetric gauge theories \cite{SW}, an important line of research
is to try to obtain microscopic derivations, from first principles, of
the proposed solutions. In the case of $\nn=2$ supersymmetry, one
needs to compute instanton contributions for any value of the
topological charge and then to sum up the resulting infinite series.
Carrying off this project required many years of developments in
instanton technology \cite{insta,instb,instc}, culminating in
Nekrasov's work \cite{nekrasova,nekrasovb}. Excellent reviews exist on
the subject \cite{instrev}.

A major remaining challenge is to apply Nekrasov's technology to the
case of $\nn=1$ gauge theories. Very little work has been done in this
subject, with the notable exception of \cite{fucito}. The main goals
are, for example, to obtain a microscopic non-perturbative derivation
of the Dijkgraaf-Vafa matrix model approach \cite{DV} and of the
generalized Konishi anomaly equations \cite{CDSW}. Our aim in the
present note is to make the first step in this direction, by
explaining in details how and why an instanton analysis can lead to a
full microscopic derivation of exact results in $\nn=1$ gauge
theories, in spite of the fact that typical vacua are strongly
coupled. We are going to derive a microscopic quantum superpotential
$\wmic$ which has two fundamental properties. First, it can be
computed exactly in the instanton approximation, and thus Nekrasov's
technology does apply. Second, the solution of the variational problem
$\d\wmic=0$ yields \emph{all} the quantum vacua of the $\nn=1$ theory,
including the strongly coupled confining vacua. Of course, at any
finite order in the instanton expansion, $\wmic$ can only describe the
vacua that can be made arbitrarily weakly coupled by adjusting the
parameters. The unbroken gauge group in these vacua has only $\u$
factors. However, if we use the exact formula for $\wmic$, then we
find all the other vacua as well, with non-abelian unbroken gauge
groups.

We focus on the well-studied example of the $\nn=1$ theory with $\uN$
gauge group, an adjoint chiral superfield $X$ and an arbitrary
polynomial tree-level superpotential $\Tr W(X)$, with
\be\label{Wdef} W'(z) = \sum_{k=0}^{d} g_{k}z^{k} =
g_{d}\prod_{i=1}^{d}(z-w_{i})\, .\ee
In this theory, the classical vacua are labeled as $|N_{i}\rangle$,
with unbroken gauge group
$\text{U}(N_{1})\times\cdots\times\text{U}(N_{d})$. The integer
$N_{i}$ is equal to the number of eigenvalues of $X$ that are equal to
$w_{i}$. This is the simplest non-trivial example for the
Dijkgraaf-Vafa theory \cite{DV}, and it displays all the essential
features of the problem. It is straightforward to generalize our
analysis to other cases.

The plan of the paper is as follows. In Section 2, we briefly discuss
different types of quantum effective superpotentials, in order to
emphasize the special conceptual r\^ole played by $\wmic$. In Section
3, we present the derivation of $\wmic$. In Section 4, we study the
stationary points of $\wmic$ and show that the set of solutions
coincide with the full set of quantum vacua of the theory. This
provides a full microscopic derivation of the gauge theory expectation
values $\langle\Tr X^{k}\rangle$ in any vacuum of the theory, and they
coincide with the Dijkgraaf-Vafa prediction. We then conclude and
explain future directions of research in Section 5.

The contribution of the present paper is mainly to set-up the right
conceptual framework to study the $\nn=1$ theories from the
microscopic point of view. A very important aspect that we do not
address is the calculation of the generalized glueball correlators
$\langle\Tr W^{\alpha}W_{\alpha}X^{k}\rangle$, where $W^{\alpha}$ is
the chiral vector superfield. These correlators play a central r\^ole
in $\nn=1$ gauge theories and in generalized anomaly equations
\cite{CDSW}. Their study from the microscopic point of view is very
interesting but technically more involved, and a detailed discussion
will appear in forthcoming papers \cite{mic2,mic3}.

\section{On quantum effective superpotentials}
\setcounter{equation}{0}

The study of quantum effective superpotentials is an extremely useful
point of view in $\nn=1$ gauge theories. There are different types of
effective superpotentials one may wish to use, and it is important to
understand the technical and conceptual differences between them. We
give a brief review of this subject in the present Section, in order
to put into perspective the properties of the microscopic
superpotential $\wmic$.

\subsection{On-shell effective superpotential}

A central object is the quantum effective superpotential
$\wl^{|0\rangle}$, defined by performing the path integral in a given
supersymmetric vacuum $|0\rangle$,
\be\label{wlowdef} e^{i\int\!\d^{4}x\left( 2N\re\int\!\d^{2}\theta\,
\wl^{|0\rangle} (\g,q) + D\text{-terms}\right)} =
\int_{|0\rangle} \! \d\mu\, e^{i{\cal S}} \, . \ee
In the above formula, $\d\mu$ denotes the path integral measure
(including the ghosts), $\mathcal S$ is the super Yang-Mills action,
$\g$ denotes collectively the couplings $g_{k}$ in the tree-level
superpotential \eqref{Wdef}, and $q$ is the instanton factor,
\be\label{qdef} q = \La^{2N}\, .\ee
The couplings $\g$ and $q$ have been promoted to arbitrary background
chiral superfields. The main property of $\wl$ is to yield the
on-shell expectation values of the chiral operators by taking the
derivative with respect to the coupling constants. If we introduce the
operators $u_{k}$ and glueball superfield $S$ defined by
\be\label{defop} u_{k} = \Tr X^{k}\, ,\quad S =
-\frac{1}{16\pi^{2}N}\Tr W^{\alpha}W_{\alpha}\, ,\ee
we have 
\be\label{vevfor}\langle 0|u_{k}|0\rangle = k
\frac{\partial\wl^{|0\rangle}}{\partial g_{k-1}}\, \cvp\quad \langle
0|S|0\rangle = q\frac{\partial\wl^{|0\rangle}}{\partial q}\,\cdotp\ee
The quantum superpotential $\wl^{|0\rangle}$ is a fundamentally
\emph{on-shell} quantity and it depends strongly on the particular
vacuum in which it is computed. To be more precise, $\wl$ is
generically a multi-valued function of the microscopic couplings $\g$
and $q$, which means that it can describe several vacua at the same
time. For example, if $W(z) = \frac{1}{2} m z^{2}$, the theory is
essentially equivalent to the pure $\nn=1$ gauge theory (after
integrating out $X$). It is well-known that this theory has $N$ vacua,
labeled as $|k\rangle$ for $0\leq k\leq N-1$, and
\be\label{wlowNpure} \wl^{|k\rangle} = Nm\, q^{1/N}e^{2i\pi k/N}\, .\ee
By doing the analytic continuations $q\rightarrow qe^{2i\pi}$, we can
smoothly interpolate between all the vacua $|k\rangle$ for any $k$.
This is possible because all these vacua are in the same confining
phase. It is then more natural to describe the physics in terms of a
single multi-valued superpotential $\wl^{|\text C)}= Nmq^{1/N}$
describing the confining phase $|\text C)$, instead of using the $N$
possible values \eqref{wlowNpure}. More generally, when the gauge
theory can be realized in several phases, we can associate a
multivalued superpotential $\wl^{|\varphi)}$ for each phase
$|\varphi)$. The degree of $\wl^{|\varphi)}$ is equal to the number of
vacua in the phase $|\varphi)$, and we can interpolate between these
vacua by doing analytic continuations. Examples have been studied in
\cite{phases}.

A particularly interesting feature of the analytic continuations is
that, in some examples, they can connect weakly coupled and strongly
coupled vacua to each other. This typically happens when fundamental
flavors are introduced in the theory. In this case, there is no
fundamental distinction between the Higgs and the confining regime
(they correspond to the same phase of the theory), and it is possible
to interpolate between the Higgs and the confining vacua \cite{FS}. In
the Higgs regime, the theory is arbitrarily weakly coupled, and thus
an instanton calculation is exact. The analytic continuation then
allows to derive exact results in the strongly coupled confining
regime, where a direct instanton analysis is not correct (and in
particular the small $q$ expansion involves fractional powers of $q$).
This is essentially the philosophy that was used long ago by Shifman
and Vainshtein to derive the gluino condensate in pure $\nn=1$
\cite{SV}, and it is at the basis of a large fraction of our
understanding of $\nn=1$ gauge theories.

So instantons can be used in some cases to derive exact results in
strongly coupled vacua. We want to know if this idea can be pushed
further: is it \emph{always} possible to analyse arbitrary $\nn=1$
vacua starting from an instanton analysis? This is clearly a necessary
condition to apply Nekrasov's technology to $\nn=1$ in general and to
provide a microscopic derivation of the exact results for this class
of theories.

The main drawback of the analysis using $\wl$ is that only vacua in
the same phase can be connected to each other. The problem clearly
comes from the fact that $\wl$ is an on-shell quantity. On the other
hand, a genuine microscopic quantum superpotential, that can describe
all the quantum vacua at the same time, must be an off-shell object.
So we need to construct off-shell quantum superpotentials.

\subsection{Integrating in}

A well-known and very easy way to do that is to ``integrate in'' some 
fields starting from $\wl$, which amounts to performing a Legendre
transform with respect to the couplings. For example, the glueball
superpotential, which plays a prominent r\^ole in the Dijkgraaf-Vafa
approach, is defined as follows. First solve the second equation in
\eqref{vevfor} to express $q$ as a function $q=\hat q(S)$ of $S$. Then
define
\be\label{gluepotdef} W_{\text{glue}}^{|0\rangle}(s;\g,q) =
\wl^{|0\rangle}\bigl(\g,\hat q(s)\bigr) + \bigl(\ln q - \ln\hat
q(s)\bigr) s\, .\ee
The superpotential $W_{\text{glue}}(s)$ is an off-shell quantity
because the variable $s$ is arbitrary and not necessarily equal to the
expectation value of the operator $S$. By construction, this
expectation value in the vacuum $|0\rangle$ can be obtained by solving
the ``quantum equations of motion''
\be\label{qem}\frac{\partial W_{\text{glue}}^{|0\rangle}}{\partial
s}\bigl(s =\langle 0|S|0\rangle\bigr) = 0\ee
and we have
\be\label{wlIO} \wl^{|0\rangle} =
W_{\text{glue}}^{|0\rangle}\bigl(s=\langle 0|S|0\rangle\bigr)\, .\ee
A priori, $W_{\text{glue}}$ depends on a vacuum $|0\rangle$, but it is
easy to see that the equation \eqref{qem} actually has several
solutions corresponding to different vacua of the same phase. For
example, in the case of \eqref{wlowNpure}, the glueball superpotential
is the Veneziano-Yankielowicz superpotential
\be\label{VYex} W_{\text{glue}}(S) =
S\ln\Bigl[q\Bigl(\frac{em}{S}\Bigr)^{N}\Bigr]\ee
for which \eqref{qem} and \eqref{wlIO} yields all the vacua
$|k\rangle$ for any $k$. In the case of the theory \eqref{Wdef}, the
vacua are labeled by the rank $r$ of the low energy gauge group. For a
given rank, the unbroken gauge group is of the form $\text
U(N_{1})\times\cdots\times\text U(N_{r})$, and the corresponding
classical vacua are of the form
$|N_{1},\ldots,N_{r},0,\ldots,0\rangle$. The glueball superpotential
can be generalized in such vacua to a function of $r$ variables
$s_{i}$ corresponding to the glueball fields of each unbroken factor
of the gauge group \cite{DV,CDSW}. It is well known that this
generalized glueball superpotential describes all the quantum vacua of
a given rank. This is interesting because there can be distinct phases
of the theory at fixed $r$. Going off-shell has thus enabled to
describe distinct phases with a unique superpotential, albeit for a
fixed value of $r$.

Another possibility is to integrate in the fields $u_{k}$ defined in
\eqref{defop}. The resulting superpotential $W_{\text{SC}}$ has been
used in the literature in the context of the ``strong coupling
approach'' to $\nn=1$, see for example \cite{CIV}. It has the same
qualitative features as the glueball superpotential. It is defined for
fixed values of the rank $r$, in which case $r$ fields (for example
$u_{1},\ldots,u_{r}$) are integrated in. This constraint comes from
the fact that at rank $r$, only $r$ of the $u_{k}$ are independent,
and thus the Legendre transform of $\wl$ is well-defined only with
respect to $r$ (or less) couplings $g_{k}$. The quantum equations of
motion
\be\label{qem2}\frac{\partial W_{\text{SC}}}{\partial u_{k}} = 0\,
,\quad 1\leq k\leq r\, ,\ee
can then be shown to describe all the quantum vacua at fixed $r$, in a
way that is equivalent to the description in terms of the glueball
superpotential \cite{CIV}.

So the superpotentials obtained by the integrating in procedure, like
$W_{\text{glue}}$ or $W_{\text{SC}}$, have nice off-shell features
(they can describe several phases at the same time), but they are not
good enough for our purposes. First, they describe vacua at fixed
values of $r$ only, and second it is only in the case $r=N$ (the
Coulomb vacuum, which can be made arbitrarily weakly coupled) that
they can be computed using an instanton analysis. We are now going to
propose a genuine microscopic off-shell superpotential, inspired by
Nekrasov's approach, that will not have these drawbacks.

\subsection{Microscopic off-shell superpotential}

Instead of picking a given vacuum as in \eqref{wlowdef}, we consider
the euclidean path integral \emph{with arbitrary boundary conditions
at infinity} for the chiral adjoint superfield $X$,
\be\label{bcdef} X_{\infty} = \diag (a_{1},\ldots,a_{N}) = \diag\a\,
.\ee
The eigenvalues $a_{i}$ can be viewed as external chiral superfields
on which the path integral depends. We shall use the notation $a_{i}$ 
(or $\a$, to denote collectively all the $a_{i}$s) either for the
chiral superfield or for its lowest, scalar, component. The
microscopic quantum effective superpotential is then defined by
\be\label{wmicdef} e^{-\int\!\d^{4}x\left( 2N\re\int\!\d^{2}\theta\,
\wmic (\a;\g,q) + D\text{-terms}\right)} = \int_{X_{\infty} = \diag\a}
\! \d\mu\, e^{-{\cal S}_{\text E}} \, . \ee
We are using explicitly the euclidean path integral, and ${\mathcal
S}_{\text E}$ is the euclidean super Yang-Mills action. By
$X_{\infty}$, we mean the value of $X$ on the three-sphere at infinity
in four dimensional euclidean space.

Several comments on the formula \eqref{wmicdef} are in order. First of
all, to be well-defined, we need to introduce an ultraviolet
regulator. Since we are going to deal with instantons, it is
convenient to use the non-commutative deformation of the theory in
order to resolve the UV singularities of the instanton moduli space.
The chiral observables we are interested in actually do not depend on
the non-commutative deformation parameter, which is real, but
introducing a non-zero deformation is necessary to obtain well-defined
integrals over the moduli space of instantons, with unambiguous
definitions of the chiral operators like the $u_{k}$ in \eqref{defop}
for any $k$. We also need to introduce an infrared regulator, to
cut-off the infrared divergence from the integration over space. We
use (implicitly) the subtle infrared regulator introduced by Nekrasov
\cite{nekrasova}, which is equivalent to turning on some particular
supergravity background (the so-called $\Omega$-background).

The reader might wonder why the path integral \eqref{wmicdef} can
depend non-trivially on the boundary conditions $\a$ when the infrared
regulator is removed. Na\"\i vely, one would expect \eqref{wmicdef} to
be projected on \eqref{wlowdef}, or on a linear combination of
contributions corresponding to different vacua. The reason why this
does not occur in the supersymmetric theories is that the $F$-term
sector is topological \cite{wittopN1}, and thus ``long distance'' can
always be pulled to ``short distance'' by rescaling the metric. The
facts that chiral correlators do not depend on the space-time
insertion points, and that the integral over the instanton moduli
space can be localized on point-like instantons, are other facets of
this property.

So we have a natural definition \eqref{wmicdef} for an off-shell
microscopic superpotential. Clearly, when $|a_{i}-a_{j}|\gg\La$, we
can compute $\wmic$ in a semiclassical approximation. Since the
corresponding instanton series has a finite radius of convergence, the
semiclassical approximation is actually exact, and thus $\wmic (\a)$
for arbitrary $\a$ can be obtained from the instanton calculation by
analytic continuation.

We now need to understand how to compute $\wmic$, and then to show
that the solutions to the equations
\be\label{qemmic}\frac{\partial\wmic}{\partial a_{i}} = 0\ee
are in one-to-one correspondence with the full set of quantum vacua of
the theory (in particular, that these equations describe the vacua for
all the possible ranks $r$).

\section{Derivation of $\wmic$}
\setcounter{equation}{0}
\begin{figure}
\centerline{\epsfig{file=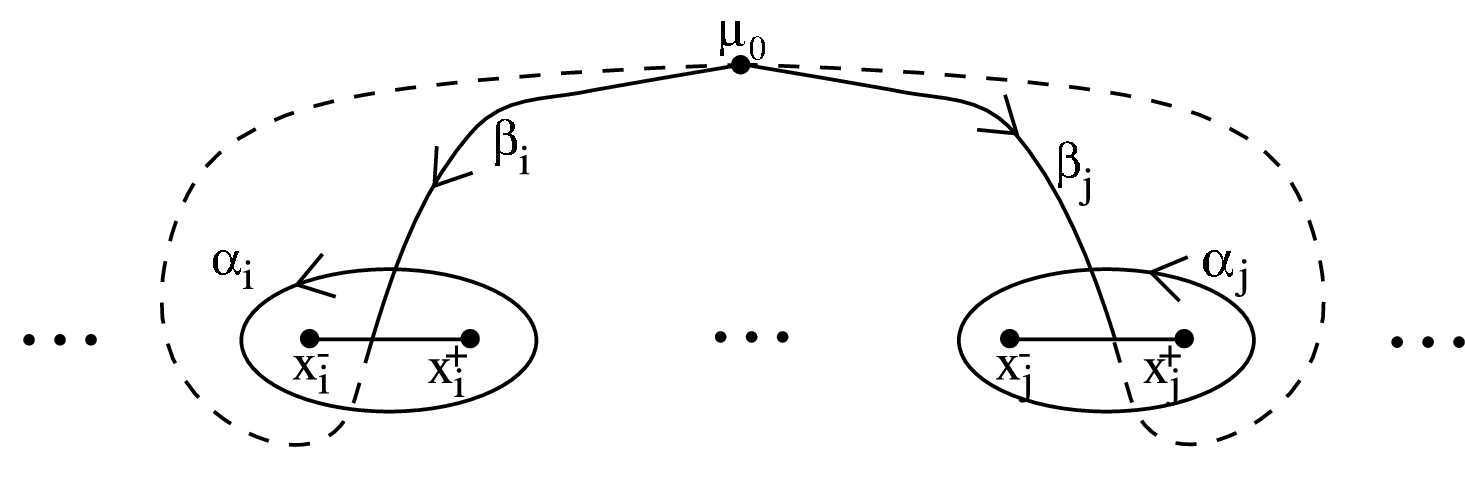,width=14cm}}
\caption{The hyperelliptic curve $\mathcal C$ defined by \eqref{SWcurve}
with the contours $\alpha_{i}$ and $\beta_{i}$ used in the main text.
The open contours $\beta_{i}$ go from the point at infinity $\mu_{0}$
on the first sheet to the point at infinity $\hat\mu_{0}$ on the
second sheet.
\label{fig}}
\end{figure}
\subsection{Off-shell correlators and $\wmic$}
\label{offshell}

From the definition \eqref{wmicdef}, it follows that
\begin{align}\label{derwm1}k\frac{\partial\wmic}{\partial g_{k-1}}
&=\vevab{\Tr X^{k}} = u_{k}(\a,\g,q)\, ,\\ \label{derwm2}
q\frac{\partial\wmic}{\partial q} &=\vevab{S} =
S(\a,\g,q)\, ,\end{align}
where $u_{k}(\a,\g,q)$ and $S(\a,\g,q)$ are the off-shell expectation
values of the operators \eqref{defop} for arbitrary values of the
boundary conditions $\a$. These functions can be computed using the
results of \cite{nekrasova} and \cite{fucito} as follows.

First, it is shown in \cite{fucito} (equation (2.17)) that
$u_{k}(\a;\g,q)$ actually does not depend on $\g$,
\be\label{ukg}u_{k}(\a,\g,q)=u_{k}(\a,q)\, .\ee
This result is a direct consequence of the localization techniques
applied to the integrals over the instanton moduli space. Using
\eqref{ukg} and \eqref{derwm1}, we deduce that
\be\label{wmicfor1}\wmic(\a,\g,q) =
\sum_{k=0}^{d}g_{k}\frac{u_{k+1}(\a,q)}{k+1} + f(\a,q) =
\bigl\langle\a\big|\Tr W(X)\big|\a\bigr\rangle + f(\a,q)\, ,\ee
where $f$ is an unknown function of $\a$ and $q$ that does not depend 
on the couplings $\g$.

When $\g=0$, the model reduces to the $\nn=2$ gauge theory, and we can
use the results of \cite{nekrasova}. Let us introduce the
Seiberg-Witten curve
\be\label{SWcurve} \mathcal C:\ y^{2} =P(z)^{2} - 4 q =
\prod_{i=1}^{N}(z-x_{i})^{2} - 4 q =
\prod_{i=1}^{N}(z-x_{i}^{-})(z-x_{i}^{+})\, ,\ee
where
\be\label{ppmdef} P_{\pm}(z) = P(z) \mp 2q^{1/2} =
\prod_{i=1}^{N}(z-x_{i}^{\pm})\, .\ee
The curve \eqref{SWcurve} is hyperelliptic of genus $N-1$. Various
contours and marked points on the curve that we use later in the text
are depicted in Figure 1. The generating function for the $u_{k}$,
\be\label{defR} R(z;\a,q) = \sum_{k\geq
0}\frac{u_{k}(\a,q)}{z^{k+1}}\,\cvp\ee
is given by \cite{nekrasova,nekrasovb}
\be\label{RNek} R(z;\a,q) = \frac{P'(z)}{\sqrt{P(z)^{2} - 4
q}}\,\cvp\ee
where the parameters $x_{i}$ entering the curve \eqref{SWcurve} are
determined in terms of the $a_{i}$s by the equations
\be\label{adef} a_{i} = \frac{1}{2i\pi}\oint_{\alpha_{i}} zR(z)\,\d z\,
.\ee
Equations \eqref{defR}, \eqref{RNek} and \eqref{adef} together with
\eqref{wmicfor1} thus determine $\wmic$ up to the function $f(\a,q)$.

Let us note that the $a_{i}$ defined by \eqref{adef} have been used in
many instances in the literature, because they are the natural
variables entering into the low energy $\nn=2$ effective action
\cite{SW}. In particular, they have simple transformation properties
under the abelian electric-magnetic duality that plays a central
r\^ole on the $\nn=2$ moduli space. However, presently, we use these
variables in a different context. For us, their relevant property is
that they precisely coincide with the boundary conditions at infinity
for the scalar field $X$. This is a highly non-trivial result that
follows from the explicit all-order instanton calculations of
\cite{nekrasova,nekrasovb}.

The $q$-dependence in $f$ can be determined by using \eqref{derwm2}. A
general formula for $S(\a,\g,q)$ has not appeared in the literature,
but it can be easily deduced from the analysis of \cite{fucito}. We do
not wish to enter into too much details here, because the analysis of
glueball operators $\Tr W^{\alpha}W_{\alpha}X^{k}$ will be presented
elsewhere \cite{mic2,mic3}. However, the case of the operator
$S\sim\Tr W^{\alpha}W_{\alpha}$ is particularly simple. The basic
formula for $S$ is
\be\label{Sform} S(\a,\g,q) =
\frac{1}{2\epsilon^{2}}\Bigl(\bigl\langle\a\big|\Tr X^{2}\Tr
W(X)\big|\a\bigr\rangle_{\epsilon} - \bigl\langle\a\big|\Tr
X^{2}\big|\a\bigr\rangle_{\epsilon}\bigl\langle\a\big|\Tr
W(X)\big|\a\bigr\rangle_{\epsilon}\Bigr)\, ,\ee
where the limit $\epsilon\rightarrow 0$ is understood. The expectation
values $\langle\cdots\rangle_{\epsilon}$ are taken for a non-zero
$\Omega$-background, the parameter $\epsilon$ measuring the strength
of this background.\footnote{The notation $\hbar$ instead of
$\epsilon$ is often used in the literature, but we find this rather
confusing in particular because $\epsilon$ is naturally a complex
parameter.} The form of the formula \eqref{Sform} shows that, to get
the glueball operator, the correlators must be computed in the
$\Omega$-background including the corrections of order $\epsilon^{2}$.
This is the basic difficulty associated with the glueball operators
and also the reason why the analysis of \cite{nekrasovb}, which is
limited to the leading order in $\epsilon$, cannot be used
straightforwardly. However, in the case of \eqref{Sform}, there is a
huge simplification due to the fact that
\be\label{simple} \frac{1}{2\epsilon^{2}}\Bigl(\bigl\langle\a\big|\Tr
X^{2}\Tr X^{k}\big|\a\bigr\rangle_{\epsilon} - \bigl\langle\a\big|\Tr
X^{2}\big|\a\bigr\rangle_{\epsilon}\bigl\langle\a\big|\Tr
X^{k}\big|\a\bigr\rangle_{\epsilon}\Bigr) =
q\frac{\partial\langle\a|\Tr X^{k}|\a\rangle_{\epsilon}}{\partial
q}\,\cvp\ee
for any $k\geq 0$. This equation was derived in \cite{fucito} and is
actually valid for any finite value of $\epsilon$. To give a hint of
the origin of \eqref{simple}, let us note that the simplifications
that allow to derive such an elegant formula are very similar to the
ones used in the all-order derivation of the Matone's relations
\cite{matone} for the $\nn=2$ prepotential \cite{fucitomatone}.
Plugging \eqref{simple} into \eqref{Sform}, we immediately obtain
\be\label{derqeq}S(\a,\g,q) = q\frac{\partial\langle\a|\Tr
W(X)|\a\rangle_{\epsilon}}{\partial q}\,\cdotp\ee
Using \eqref{wmicfor1} and \eqref{derwm2}, we see that $f(\a,q)=f(\a)$
can depend only on $\a$. We can thus determine $f$ by looking at the
classical limit $q\rightarrow 0$ for which it is clear that $f=0$.
Note that the classical limit is perfectly smooth since $\wmic$ is
given by an instanton expansion (this is unlike the classical limit
for the glueball superpotential for example; for this reason,
$W_{\text{glue}}$ can only be determined up to an arbitrary function
of the glueball fields $s_{i}$ by studying the correlators
\cite{CDSW}). 

Thus we have derived the fundamental formula
\be\label{Wmicresa}\wmic(\a,\g,q) = \bigl\langle\a\big|\Tr
W(X)\big|\a\bigr\rangle\, .\ee
Using \eqref{defR} and \eqref{RNek}, this is equivalent to
\be\label{Wmicresb} \wmic(\a,\g,q) =
\frac{1}{2i\pi}\oint_{\alpha}W(z)R(z;\a,q)\,\d z=
\frac{1}{2i\pi}\oint_{\alpha}\frac{W(z)P'(z)}{\sqrt{P(z)^{2}-4q}}\,\d
z \, \cvp\ee
where the contour $\alpha = \sum_{i=1}^{N}\alpha_{i}$.

\subsection{Using the $\uR$ symmetry}

As emphasized in \cite{ferproof} in the case of the glueball
superpotential, R-symmetries put strong constraints on the effective
superpotential. We can actually rederive \eqref{Wmicresa} by using the
$\uR$ symmetry of our model. The charges of the superspace coordinates
$\theta^{\alpha}$, instanton factor $q$, chiral superfield $X$, vector
superfield $W^{\alpha}$, boundary conditions $\a$, couplings $\g$ and
superpotential $\wmic$ are given in the following table,
\be\label{charges}
\begin{matrix}
& \theta^{\alpha} & q & X & W^{\alpha} & \a & \g & \wmic \\
\uR & 1 & 0 & 0 & 1 & 0 & 2 &\hphantom{,\,} 2 \, .
\end{matrix}\ee
Performing an infinitesimal $\uR$ transformation in the path integral
\eqref{wmicdef}, we obtain
\be\label{uRconst} 2\wmic = \sum_{k\geq 0}
2g_{k}\frac{\partial\wmic}{\partial g_{k}}\, \cvp\ee
and using \eqref{derwm1} we find \eqref{Wmicresa} again.

\subsection{Relation with the strong coupling approach}

The $\uR$ symmetry can be used to constrain the various types of
effective superpotentials discussed previously. For example, it yields
\be\label{uRc2} \wl^{|0\rangle} = \bigl\langle 0\big|\Tr
W(X)\big|0\bigl\rangle\, , \ee
and similarly a very useful constraint is obtained for the
Dijkgraaf-Vafa glueball superpotential as explained in
\cite{ferproof}. Let us look in more details at the superpotentials
$\wsc$ used in the strong coupling approach. We denote by
$\wsc^{(r)}(u_{1},\ldots,u_{r})$ the superpotential relevant to the
vacua of rank $r$. The fact that the variables $u_{k}$ have $\uR$
charge zero makes the $\wsc^{(r)}$ somewhat similar to $\wmic$ in the
sense that the $\uR$ symmetry also implies that
\be\label{wscuR}\wsc^{(r)} = \sum_{k\geq 0}
g_{k}\frac{\partial\wsc^{(r)}}{\partial g_{k}}=\bigl\lfloor r\big|\Tr
W(X)\big|r\bigr\rfloor \, .\ee
By $\lfloor r|\Tr W(X)|r\rfloor$, we mean that the expectation value
is computed by taking into account the constraints that correspond to
being in a vacuum of rank $r$. Explicitly, the $u_{k'}$ for $k'>r$ are
functions of the $u_{k}$ for $1\leq k\leq r$, and thus we have a
formula of the form
\be\label{rankrexp} \wsc^{(r)}(u_{1},\ldots,u_{r}) =
\sum_{k=1}^{r}g_{k}\frac{u_{k+1}}{k+1} +
\sum_{k=r+1}^{d+1}g_{k}\frac{u_{k+1}(u_{1},\ldots,u_{r};q)}{k+1}\,\cdotp
\ee
At the classical level, it is straightforward to write down the
constraints that define implicitly the functions
$u_{k}(u_{1},\ldots,u_{r};q=0)$. For example, in the simplest $r=1$
case for which the matrix $X$ is proportional to the identity, we have
$u_{k}=N^{1-k}u_{1}^{k}$. The main drawback of the strong coupling
approach is that the constraints are not known a priori at the quantum
level. They must be postulated based on some physical insights. The
correct guess, that originates from \cite{SW}, is that the rank $r$
vacua are characterized by the factorization condition
\be\label{fact1} P(z)^{2} - 4 q = H_{N-r}(z)^{2}R_{2r}(z)\, ,\ee
where $H_{N-r}$ and $R_{2r}$ are polynomials of degrees $N-r$ and $2r$
respectively. This condition is equivalent to the fact that the curve
\eqref{SWcurve} degenerates to a genus $r-1$ surface. Physically, the
$\nn=2$ theory then has $N-r$ massless monopoles which can condense
when $W$ is turned on, higgsing the low energy gauge group from
$\u^{N}$ to $\u^{r}$.

Let us assume that $d= N$.\footnote{We could assume more generally
that $d\geq N$, but this does not bring any new interesting insight.}
Then there exists a rank $r=N$ vacuum, the Coulomb vacuum,
corresponding to the unbroken gauge group $\u^{N}$, described by the
superpotential $\wsc^{(N)}$. In the rank $N$ case, the condition
\eqref{fact1} is trivially satisfied: the variables
$u_{1},\ldots,u_{N}$ are independent. It is convenient to use the set
of variables $\x=(x_{1},\ldots,x_{N})$ which, according to
\eqref{RNek}, are related to the $u_{k}$ for small enough values of
$k$ by
\be\label{xuk} u_{k} = \sum_{i=1}^{N}x_{i}^{k}\, ,\quad 1\leq k\leq
2N-1\, .\ee
Equations \eqref{rankrexp} and \eqref{RNek} then yield
\be\label{wscN}\wsc^{(N)}(\x) =\sum_{i=1}^{N} W(x_{i}) =
\frac{1}{2i\pi}\oint_{\alpha}\frac{W(z)P'(z)}{\sqrt{P(z)^{2}-4q}}\,\d
z \, . \ee
Comparing with \eqref{Wmicresb}, we see that
\be\label{wscwmicrel}\wmic(\a) = \wsc^{(N)}(\x)\, .\ee
This may look like a rather surprising formula, in view of the
important conceptual differences between $\wmic$ and $\wsc^{(N)}$. In
particular, we have advertised that the solutions to the equations
\eqref{qemmic} are all physical and describe the full set of vacua of
the quantum theory. On the other hand, the equations
\be\label{qemSCbis}\frac{\partial\wsc^{(N)}}{\partial x_{i}} =
W'(x_{i}) = 0\ee
describe a single vacuum, the weakly coupled Coulomb vacuum. This
corresponds to the solution of \eqref{qemSCbis} for which all the
$x_{i}$ are distinct and equal to the classical values, $x_{i}=w_{i}$
up to permutations. The solutions of \eqref{qemSCbis} for which some
of the $x_{i}$ coincide are not physical. This is a trivial artefact
of the variables $\x$. To be fully rigorous, we should follow the
prescription from the integrating in procedure and use instead the
variables $u_{1},\ldots,u_{N}$ \eqref{xuk}. The equations
\be\label{qemSCtris}\frac{\partial\wsc^{(N)}}{\partial u_{k}} = 0\ee
then have only one solution corresponding to \eqref{qemSCbis} with all
the $x_{i}$ distinct. 

The fact that the set of stationary points strongly depends on the
variables we use is at the heart of the fundamental difference between
$\wmic$ and $\wscN$. There is a lot of physics in the choice of the
variables, $\x$ or $\a$. This is one of the main point of the present
paper. The variables $\x$ enter when one considers the integrating in
procedure, as in \cite{CIV}, because of the relation \eqref{xuk}. On
the other hand, a microscopic point of view singles out the variables
$\a$, as explained in \ref{offshell}.

\section{The stationary points of $\wmic$}
\setcounter{equation}{0}

We are now going to solve the equations \eqref{qemmic} and prove the
claims made earlier in the paper. We use a strategy based on the
relationship between $\wsc$ and $\wmic$. This has the advantage of
exhibiting clearly the differences between the usual integrating in
approach and the present microscopic approach. Another derivation of
the same results is also possible using generalized Riemann bilinear
relations. It will be presented in a forthcoming paper \cite{mic3}.

Let us use \eqref{wscwmicrel} to rewrite \eqref{qemmic} as
\be\label{cov1} \frac{\partial\wmic}{\partial a_{i}} =
\sum_{i=1}^{N}A_{ij} \frac{\partial\wscN}{\partial x_{j}}=0\,\cvp\ee
where we have introduced the matrix
\be\label{Adef} A_{ij} = \frac{\partial x_{j}}{\partial
a_{i}}\,\cdotp\ee
The relation between the variables $\x$ and $\a$ is given explicitly
by \eqref{adef} and \eqref{RNek}. The equations \eqref{cov1} can be
solved in \emph{two} ways:\smallskip\\
$\bullet$ Equation \eqref{qemSCbis} is satisfied. This case
corresponds to the Coulomb vacuum as discussed above.\\
$\bullet$ Equation \eqref{qemSCbis} is \emph{not} satisfied, but
$\partial\wscN/\partial x_{i}$ is an eigenvector of $A$ of zero
eigenvalue. This is possible only if the rank of the matrix $A$ is
$r<N$. We are going to show that these solutions correspond precisely
to the vacua of rank $r<N$.

\subsection{Mathematical preliminaries}

\noindent\textsc{One-forms} $h_{i}$: Let us introduce the differential
forms on the curve $\mathcal C$ \eqref{SWcurve}
\be\label{hidef} h_{i} = \psi_{i}(z)\,\d z = \frac{p_{i}}{y}\,\d z\,
,\ee
where the $p_{i}(z) = z^{N-1}+\cdots$ are monic polynomials of degree
$N-1$ fixed by the conditions
\be\label{hicond} \frac{1}{2i\pi}\oint_{\alpha_{i}}h_{j} =
\delta_{ij}\, .\ee
When the curve \eqref{SWcurve} is of genus $N-1$ (i.e.\ it is not
degenerate), the $h_{i}$s form a canonical basis of the vector space
$\mathscr L^{(N)}$ defined by the following constraint on the divisor
of one-forms on $\mathcal C$,
\be\label{Ldef}\mathscr L^{(N)} = \{\text{one-forms}\ \eta\mid
(\eta)+\mu_{0} + \hat\mu_{0}\geq 0 \} \, .\ee
This corresponds to one-forms that are holomorphic except possibly at
infinity on either sheet where they may have a simple pole. Note that
the fact that the $h_{i}$s are linearly independent follows from
\eqref{hicond} and the fact that they generate $\mathscr L^{(N)}$ is a
straightforward consequence of the Riemann-Roch theorem.

It is useful to understand the one-forms $h_{i}$ also in the case of a
degenerate curve of the form \eqref{fact1}. Let us study what happens
when two branch cuts join together, for example the branch cuts
encircled by the contours $\alpha_{1}$ and $\alpha_{2}$. The genus of
the curve then drops from $N-1$ to $N-2$. In the notation of
\eqref{SWcurve} and \eqref{ppmdef}, this corresponds to
$x_{1}^{+}=x_{2}^{+}$ or $x_{1}^{-}=x_{2}^{-}$ (we cannot have
$x_{1}^{\pm}=x_{2}^{\mp}$ because $P_{+}$ and $P_{-}$ do not have
common roots). Let us choose for example $x_{1}^{+}=x_{2}^{+}=b_{1}$,
and
\be\label{degcurve} y^{2} = (z-b_{1})^{2}R_{2N-2}(z)\, .\ee
Naively, the one-forms
\be\label{hideg} h_{i} =
\frac{p_{i}}{(z-b_{1})\sqrt{R_{2N-2}}}\,\d z\ee
then have poles at $z=b_{1}$ on the first and second sheets. However,
this does not happen, because $p_{j}(b_{1}) = 0$ for all $j$. This
follows from the constraints \eqref{hicond} for $i=1$ or $i=2$.
Indeed, if we had $p_{j}(b_{1})\not = 0$, then the contour integrals
would have a logarithmic divergence in the degenerate limit. So we see
that the $h_{i}$ remains holomorphic at finite $z$ on \eqref{hideg},
with simple poles at infinity. In other words, the $h_{i}$ belongs to
the space $\mathscr L^{(N-1)}$ defined as in \eqref{Ldef} but on the
curve
\be\label{gencur} y_{N-1}^{2} = R_{2N-2}(z)\, .\ee
Using \eqref{hicond}, it follows that a canonical basis
$\{h_{i}^{(N-1)}\}_{2\leq i\leq N}$ of $\mathscr L^{(N-1)}$ is given
by
\be\label{hireldeg} h_{1} = h_{2} = h_{2}^{(N-1)}\, ,\quad h_{i} =
h^{(N-1)}_{i}\quad \text{for}\ i\geq 3\, ,\ee
in the degenerate limit.

In the general case, \eqref{SWcurve} can degenerate to a genus $r-1$
curve
\be\label{yrdef} y_{r}^{2} = R_{2r}(z)\ee
with
\be\label{degcgen} y = H_{N-r}(z)\, y_{r} =
\prod_{\ell=1}^{N-r}(z-b_{\ell})\,y_{r}\, .\ee
The $h_{i}$s then generate the vector space $\mathscr L^{(r)}$ of
one-forms on \eqref{yrdef} that are holomorphic at finite $z$ with at
most simple poles at infinity, but with relations like $h_{i}=h_{j}$
depending on which cuts have joined. In particular, we have
\be\label{piconst} p_{i}(b_{\ell}) = 0\, .\ee
Clearly, the rank of the system $\{ h_{i}\}$ is given by
\be\label{rankhi} \rank \{h_{i}\}_{1\leq i\leq N} = r\, .\ee

\noindent\textsc{One-forms} $\eta_{i}$: Let us introduce another basis
of $\mathscr L^{(N)}$ given by
\be\label{etaidef}\eta_{i} = \phi_{i}(z)\,\d z =\frac{q_{i}}{y}\,\d
z\, \cvp\ee
with
\be\label{qidef} q_{i}(z) = \prod_{j\not = i} (z-x_{j}) =
-\frac{\partial P(z)}{\partial x_{i}}\,\cdotp\ee
We assume that the $x_{i}$ are all distinct.

The one-forms $\eta_{i}$ belong to $\mathscr L^{(r)}$ only for $r=N$
because of the $2(N-r)$ poles at $z=b_{\ell}$. Actually, even when the
curve degenerates, the rank of the system $\{\eta_{i}\}$ doesn't
change,
\be\label{ranketai} \rank \{\eta_{i}\}_{1\leq i\leq N} = N\, .\ee
This is a consequence of the linear independence of the polynomials
$q_{i}$. Moreover, because the $q_{i}$s form a basis for the
polynomials of degree at most $N-1$, there always exists a matrix $A$
such that\footnote{We show later that this definition is consistent
with \eqref{cov1}.}
\be\label{pqrel} p_{i}(z) = \sum_{j=1}^{N} A_{ij}\, q_{j}(z)\, .\ee
Using \eqref{hidef} and \eqref{etaidef}, we also have
\be\label{hetarel}\psi_{i}= \sum_{j=1}^{N} A_{ij}\,\phi_{j}\, ,\quad
h_{i} = \sum_{j=1}^{N}A_{ij}\,\eta_{j}\, .\ee
The important point is that these relations are always valid,
including in the cases where the curve degenerate, because the
$q_{i}$s are always linearly independent when the $x_{i}$ are all
distinct. Using \eqref{rankhi} and \eqref{ranketai}, we also deduce
that
\be\label{rankA}\rank A = r\, .\ee
Conversely, $A$ of rank $r$ clearly implies \eqref{rankhi} which
implies that the curve is of genus $r-1$.

\noindent\textsc{Variations of the Seiberg-Witten differential}:
Consider now the Seiberg-Witten differential
\be\label{SWdiff} \lsw = zR(z)\,\d z\, .\ee
An important property of $R(z)$ is that the solution to
\be\label{FRrel} \frac{F'(z)}{F(z)} = R(z)\ee
is a function $F(z)$ defined on the Seiberg-Witten curve,
\be\label{Fdef} F(z;\a,q) =
\bigl\langle\a\big|\det(z-X)\big|\a\bigr\rangle =
\frac{1}{2}\Bigl(P(z) + \sqrt{P(z)^{2} - 4q}\Bigr)\, . \ee
Another useful identity is that
\be\label{FPdelta} \frac{\delta F}{F} = \frac{\delta P}{y}\,\cvp\ee
where the variation $\delta$ is with respect to any parameter, for
example the $a_{i}$s or the $x_{i}$s. To compute $\delta\lsw$, it is
then convenient to write
\be\label{SWdiff2} \lsw = -\ln F\,\d z + \d (z\ln F)\, .\ee
We get
\be\label{SWdvar}\delta\lsw = -\frac{\delta F}{F}\,\d z + \d
\Bigl(z\frac{\delta F}{F}\Bigr)=-\frac{\delta P}{y}\,\d z + \d
\Bigl(z\frac{\delta P}{y}\Bigr)\, .\ee
Note that when the curve is non-degenerate, $-\delta P\d
z/y\in\mathscr L^{(N)}$.

We can use the above results to compute the derivatives of $\lsw$ with
respect to $a_{i}$ and to $x_{i}$. Using
\be\label{lswdid} \frac{1}{2i\pi}\oint_{\alpha_{i}}
\frac{\partial\lsw}{\partial a_{j}} = \delta_{ij}\, ,\ee
which comes from taking the derivative of \eqref{adef} with respect to
$a_{j}$, we get, in terms of \eqref{hidef},
\be\label{Pderplus}-\frac{1}{y}\frac{\partial P}{\partial a_{i}}\,\d z
= h_{i}\, ,\ee
and thus
\be\label{lswder1}\frac{\partial\lsw}{\partial a_{i}} = h_{i} - \d
(z\psi_{i})=\psi_{i}\,\d z - \d (z\psi_{i})\, .\ee
Similarly, we get in terms of \eqref{etaidef}
\be\label{lswder2} \frac{\partial\lsw}{\partial x_{i}} = \phi_{i}\,\d
z - \d (z\phi_{i})\, . \ee

\noindent\textsc{The tangent space to} $\Sigma_{r}$: Let us define
$\Sigma_{r}$ to be the $r$-dimensional surface in $\x$-space on which 
the Seiberg-Witten curve degenerates to a genus $r-1$ surface. Let us 
show that the vectors
\be\label{Sigmartg} e_{i} =\sum_{j=1}^{N}A_{ij}\frac{\partial}{\partial
x_{j}}\ee
generate the tangent space to $\Sigma_{r}$. Due to \eqref{rankA}, all 
we have to show is that $e_{i}(\x)\in T_{\x}\Sigma_{r}$. The result
is true essentially by construction, but let us see explicitly how it 
works.

The surface $\Sigma_{r}$ is defined by the equation \eqref{fact1}.
Thus an arbitrary vector $\nabla\in T_{\x}\Sigma_{r}$ if and only if
there exists polynomials $H_{N-r}$ and $R_{2r}$ such that
\be\label{condtg} \nabla\cdot\bigl( P(z)^{2} - 4 q\bigr) =
\nabla\cdot\bigl(H_{N-r}(z)^{2}R_{2r}(z)\bigr)
\ee
or equivalently
\be\label{condtg2} 2P\nabla\cdot P = H_{N-r}\bigl( 2
R_{2r}\nabla\cdot H_{N-r} + H_{N-r}\nabla\cdot
R_{2r}\bigr)\, .\ee
Equation \eqref{fact1} implies that $P$ and $H_{N-r}$ cannot have
common roots. Thus \eqref{condtg2} implies that
\begin{align}\label{ct1} \nabla\cdot P &= H_{N-r} Q_{r-1}\,
,\\ \label{ct2} 2P Q_{r-1} &= 2 R_{2r}\nabla\cdot H_{N-r} +
H_{N-r}\nabla\cdot R_{2r}\, ,\end{align}
for some degree $r-1$ polynomial $Q_{r-1}$. Conversely, assume that
the vector $\nabla$ is such that \eqref{ct1} is satisfied. Then the
equation \eqref{ct2} can be viewed as a constraint that determines the
polynomials $\nabla\cdot H_{N-r}$ and $\nabla\cdot R_{2r}$ (the
equation is of degree $N+r-1$, for $N+r$ unknown in $\nabla\cdot
H_{N-r}$ and $\nabla\cdot R_{2r}$).

So all we have to show is that the vectors \eqref{Sigmartg} satisfy
\be\label{eicons} e_{i}\cdot P =
H_{N-r}p_{i}^{(r)}=\prod_{\ell=1}^{N-r}(z-b_{\ell})p_{i}^{(r)}\ee
for some polynomials $p_{i}^{(r)}$ of degrees $r-1$. This follows
immediately from \eqref{qidef} and \eqref{pqrel}, which imply that
\be\label{eiact}e_{i}\cdot P = -p_{i}(z)\, ,\ee
and from \eqref{piconst}. Finally, we have derived that
\be\label{tgres}T_{\x}\Sigma_{r} = \Vect\bigl[ e_{i}\bigr]_{1\leq
i\leq N}\, .\ee
\subsection{Solving $\d\wmic=0$}

Let us rewrite \eqref{Wmicresb} in the form
\be\label{Wmiccal}\wmic =
\frac{1}{2i\pi}\oint_{\alpha}\frac{W(z)}{z}\,\lsw\, .\ee
Using \eqref{lswder1} and performing an integration by part, we get
\be\label{Wmicder}\frac{\partial\wmic}{\partial a_{i}} =
\frac{1}{2i\pi}\oint_{\alpha}W'(z)\psi_{i}(z)\,\d z\, .\ee
Similarly, from \eqref{wscN} and \eqref{lswder2} we obtain
\be\label{Wscder}\frac{\partial\wscN}{\partial x_{i}} = W'(x_{i})=
\frac{1}{2i\pi}\oint_{\alpha}W'(z)\phi_{i}(z)\,\d z\, .\ee

\noindent\textsc{The case of distinct} $x_{i}$: Let us assume for the
moment that the $x_{i}$ are all distinct. Then using \eqref{hetarel},
\eqref{Wmicder} and \eqref{Wscder}, we find \eqref{cov1}. This
equation is valid for any genus $r-1$ of the Seiberg-Witten curve, and
can be written in terms of the vector fields \eqref{Sigmartg} as
\be\label{eionW} e_{i}\cdot\wscN = 0\, .\ee
The most general solution is labeled by $r$, and, for a given $r$, we
find using \eqref{tgres} that it corresponds to extrema of $\wscN$ on
the surface $\Sigma_{r}$. This is exactly the prescription used in the
strong coupling approach. Using \eqref{Wmicder} and \eqref{piconst},
finding the extrema along $\Sigma_{r}$ is equivalent to imposing
\be\label{impose} \oint_{\alpha}\frac{W'(z)}{y_{r}}Q_{r-1}\,\d z =
0\ee
for any degree $r-1$ polynomial $Q_{r-1}$. Since the integral in
\eqref{impose} simply picks the simple pole at infinity, this yields
the condition
\be\label{impose2} \frac{W'}{y_{r}} = \tilde H_{d-r}+ \mathcal
O(1/z^{r+1})\,\ee
for some degree $d-r$ polynomial $\tilde H_{d-r}$, or
\be\label{impose3} W' = \tilde H_{d-r}y_{r} + \mathcal O(1/z)\,
.\ee
Taking the square of \eqref{impose3}, we find
\be\label{impose4} W'^{2} = \tilde H_{d-r}^{2}R_{2r} + \mathcal
O(z^{d-1})\, .\ee
Since both $W'^{2}$ and $\tilde H_{d-r}^{2}R_{2r}$ are polynomials,
this is equivalent to the existence of a degree $d-1$ polynomial
$\Delta_{d-1}$ such that
\be\label{imposef} W'(z)^{2} - \Delta_{d-1}(z) =
H_{d-r}(z)^{2}R_{2r}(z)\, .\ee
This is the usual factorization condition which, together with
\eqref{fact1}, yields the full solution of the theory.

\noindent\textsc{The case of} $x_{i}=x_{j}$: It might happen that, for
some particular values of the couplings, some solutions correspond to
$x_{i}=x_{j}$ for a pair of distinct indices $i$ and $j$. The previous
analysis doesn't apply immediately in this case because for
$x_{i}=x_{j}$, we have $q_{i}=q_{j}$, thus the polynomials $q_{k}$ are
no longer independent and the matrix $A$ is not well-defined. However,
let us show that the formulas behave smoothly when we approach such a
point. Let us start from a case where $x_{i}$ and $x_{j}$ are very
close to each other,
\be\label{epsdef} x_{i}-x_{j} = \epsilon\, .\ee
Then $q_{i}-q_{j}\sim\epsilon$. Using \eqref{pqrel}, we see that in
the limit $\epsilon\rightarrow 0$, the components $A_{ki}$ and
$A_{kj}$ diverge as
\be\label{Adiv} A_{ki}\sim\frac{b_{ki}}{\epsilon}\sim -A_{kj}\, .\ee
On the other hand, the potentially diverging terms in \eqref{cov1}
read
\be\label{smooth} A_{ki}\frac{\partial\wscN}{\partial x_{i}} +
A_{kj}\frac{\partial\wscN}{\partial
x_{j}}\sim\frac{b_{ki}}{\epsilon}\bigl(W'(x_{i})-W'(x_{j})\bigr)\sim
b_{ki}W''(x_{i})\, , \ee
and thus the limit $\epsilon\rightarrow 0$ is smooth.

\section{Conclusion and outlook}
\setcounter{equation}{0}

We have shown that a microscopic approach to $\nn=1$ gauge theories,
based on Nekrasov's instanton technology, is possible. The stationary
points of the microscopic superpotential $\wmic(\a)$ yield all the
quantum vacua, including the strongly coupled confining vacua, and the
result is consistent with the strong coupling approach or the
Dijkgraaf-Vafa matrix model. In particular, at the extrema of $\wmic$,
the gauge theory resolvent \eqref{defR}, \eqref{RNek} precisely
coincide with the prediction of the matrix model. In other word, we
have obtained a full microscopic description of the expectation values
$\langle\Tr X^{k}\rangle$ in all the vacua of the theory.

One of the most interesting potential application of the
superpotential $\wmic$ is the non-perturbative study of the
generalized Konishi anomaly equations. At the moment, only a
perturbative analysis of these equations has appeared \cite{CDSW},
whereas the equations are supposed to be valid at the non-perturbative
level (see for example \cite{ferchiral} for a discussion). On general 
grounds, one may expect to have relations like
\be\label{ano}\delta\wmic = \mathscr A\, ,\ee
where $\mathscr A$ is the anomaly polynomial and $\delta$ is a
suitable variation. At the perturbative level, the variations $\delta$
one must consider \cite{CDSW} act on the fields as $\delta X \sim
X^{n+1}$, $\delta X\sim W^{\alpha}W_{\alpha} X^{n+1}$, and thus
generate a sort of super Virasoro algebra. At the non-perturbative
level, the variations $\delta$ and associated algebra must be quantum
corrected (this happens because the transformations are non-linear).
The corrections can in principle be studied starting from \eqref{ano}.

To complete the above program, we need to study in full details the
glueball operators
\be\label{gluedef} v_{k}(\a,\g,q) = -\frac{1}{16\pi^{2}}\vevab{\Tr
W^{\alpha}W_{\alpha} X^{k}}\ee
or the associated generating function
\be\label{Safor} S(z;\a,\g,q) = \sum_{k\geq 0}\frac{
v_{k}(\a,\g,q)}{z^{k+1}}\, \cdotp\ee
The function $S(z;\a,\g,q)$ is not known for arbitrary values of the
boundary conditions $\a$. As sketched in \ref{offshell}, it depends on
subleading corrections in $\epsilon$ in Nekrasov's formalism.
Computing this function, showing that it enters the anomaly equations
\eqref{ano} at the non-perturbative level as derived in \cite{CDSW} in
perturbation theory, and that it coincides with the matrix model
prediction at the extrema of $\wmic$ will be a central topic in
forthcoming publications \cite{mic2,mic3}.

\subsection*{Acknowledgements}

I would like to thank Vincent Wens for useful discussions and for his
careful reading of the manuscript. This work is supported in part by
the belgian Fonds de la Recherche Fondamentale Collective (grant
2.4655.07), the belgian Institut Interuniversitaire des Sciences
Nucl\'eaires (grant 4.4505.86), the Interuniversity Attraction Poles
Programme (Belgian Science Policy) and by the European Commission FP6
programme MRTN-CT-2004-005104 (in association with V.\ U.\ Brussels).
The author is on leave of absence from Centre National de la Recherche
Scientifique, Laboratoire de Physique Th\'eorique de l'\'Ecole Normale
Sup\'erieure, Paris, France.

\renewcommand{\thesection}{\Alph{section}}
\renewcommand{\thesubsection}{\arabic{subsection}}
\renewcommand{\theequation}{A.\arabic{equation}}
\setcounter{section}{0}

\begin{thebibliography}{99}
%
\bibitem{SW}{N.~Seiberg and E.~Witten, \npb{426}{1994}{19}, erratum 
{\bf B 430} (1994) 485, hep-th/9407087; \npb{431}{1994}{484}, 
hep-th/9408099.}
%
\bibitem{insta}{N.~Dorey, V.V.~Khoze and M.P.~Mattis,
\prd{54}{1996}{2921}, hep-th/9603136,\\
F.~Fucito and G.~Travaglini, \prd{55}{1997}{1099},
hep-th/9605215,\\
N.~Dorey, V.V.~Khoze and M.P.~Mattis, \prd{54}{1996}{7832},
hep-th/9607202,\\
N.~Dorey, V.V.~Khoze and M.P.~Mattis, \plb{396}{1997}{141},
hep-th/9612231,\\
V.V.~Khoze, M.P.~Mattis and M.J.~Slater, \npb{536}{1998}{69},
hep-th/9804009.}
%
\bibitem{instb}{D.~Bellisai, F.~Fucito, A.~Tanzini and G.~Travaglini, 
\plb{480}{2000}{365}, hep-th/0002110,\\
D.~Bellisai, F.~Fucito, A.~Tanzini and G.~Travaglini,
\jhep{07}{2000}{017}, hep-th/0003272,\\
N.~Dorey, T.J.~Hollowood and V.V.~Khoze, \jhep{03}{2001}{040},
hep-th/0011247,\\
F.~Fucito, J.F.~Morales and A.~Tanzini, \jhep{07}{2001}{012},
hep-th/0106061.}
%
\bibitem{instc}{T.J.~Hollowood, \jhep{03}{2002}{038},
hep-th/0201075,\\
T.J.~Hollowood, \npb{639}{2002}{66}, hep-th/0202197.}
%
\bibitem{nekrasova}{N.~Nekrasov, \atmp{7}{2004}{831}, hep-th/0206161,\\
N.~Nekrasov, {\it Seiberg-Witten Prepotential from Instanton
Counting,} Proceedings\ of the International Congress of
Mathematicians (ICM 2002), hep-th/0306211.}
%
\bibitem{nekrasovb}{N.~Nekrasov and A.~Okounkov, {\it Seiberg-Witten
Theory and Random Partitions,} hep-th/0306238,\\
N.~Nekrasov and S.~Shadchin, \cmp{252}{2004}{359}, hep-th/0404225.}
%
\bibitem{instrev}{N.~Dorey, T.J.~Hollowood, V.V.~Khoze and
M.P.~Mattis, \pr{371}{2002}{231}, hep-th/0206063,\\
M.~Bianchi, S.~Kovacs and G.~Rossi, \emph{Instantons and
Supersymmetry}, hep-th/0703142.}
%
\bibitem{fucito}{F.~Fucito, J.F.~Morales, R.~Poghossian and
A.~Tanzini, \jhep{01}{2006}{031}, hep-th/0510173}.
%
\bibitem{DV}{R.~Dijkgraaf and C.~Vafa, {\it A Perturbative Window into
Non-Perturbative Physics,} hep-th/0208048.}
%
\bibitem{CDSW}{F.~Cachazo, M.R.~Douglas, N.~Seiberg and E.~Witten, 
\jhep{12}{2002}{071}, hep-th/0211170.}
%
\bibitem{mic2}{F.~Ferrari, S.~Kuperstein and V.~Wens, \emph{Glueball
operators and the microscopic approach to $\nn=1$ gauge theories},
arXiv:0708.1410, to appear in \emph{JHEP}.}
%
\bibitem{mic3}{F.~Ferrari, \emph{Extended $\nn=1$ super Yang-Mills
theory},  arXiv:0709.0472, to appear in \emph{JHEP}.}
%
\bibitem{phases}{F.~Ferrari, \prd{67}{2003}{85013}, hep-th/0211069,\\
F.~Cachazo, N.~Seiberg and E.~Witten,
\jhep{02}{2003}{042}, hep-th/0301006,\\
F.~Ferrari, \plb{557}{2003}{290}, hep-th/0301157,\\
F.~Cachazo, N.~Seiberg and E.~Witten,
\jhep{04}{2003}{018}, hep-th/0303207.}
%
\bibitem{FS}{T.~Banks and E.~Rabinovici, \npb{160}{1979}{349},\\
E.H.~Fradkin and S.H.~Shenker, \prd{19}{1979}{3682}.}
%
\bibitem{SV}{M.~Shifman and A.~Vainshtein, \npb{296}{1988}{445}.}
%
\bibitem{CIV}{F.~Cachazo, K.~Intriligator and C.~Vafa, 
\npb{603}{2001}{3}, hep-th/0103067.}
%
\bibitem{wittopN1}{E.~Witten, \jmp{35}{1994}{5101}, hep-th/9403195.}
%
\bibitem{matone}{M.~Matone, \plb{357}{1995}{342}, hep-th/9506102,\\
J.~Sonnenschein, S.~Theisen and S.~Yankielowicz, \plb{367}{1996}{145},
hep-th/9510129.}
%
\bibitem{fucitomatone}{R.~Flume, F.~Fucito, J.F.~Morales,
R.~Poghossian, \jhep{04}{2004}{008}, hep-th/0403057.}
%
\bibitem{ferproof}{F.~Ferrari, \jhep{06}{2006}{039}, hep-th/0602249.}
%
\bibitem{ferchiral}{F.~Ferrari, \npb{770}{2007}{371}, hep-th/0701220.}
%
\end{thebibliography}
\end{document}